\documentclass[prb,twocolumn,epsf,floats,aps]{revtex4-1}

\usepackage{amsmath}
\usepackage{enumerate}
\usepackage{amssymb}
\usepackage{graphicx}
\usepackage{epsfig}
\usepackage{setspace}
\usepackage[hypertex]{hyperref}
\usepackage{subfig}
\usepackage{bbm}
\usepackage{graphics}

\begin{document}

\title{Of symmetries, symmetry classes, and symmetric spaces:\\
from disorder and quantum chaos to topological insulators}

\author{Martin R.\ Zirnbauer}
\affiliation{Institut f\"ur Theoretische Physik, Universit\"at zu K\"oln, Z\"ulpicher Stra{\ss}e 77, 50937 K\"oln, Germany}

\date{March 27, 2012}

\begin{abstract}
Quantum mechanical systems with some degree of complexity due to multiple scattering behave as if their Hamiltonians were random matrices. Such behavior, while originally surmised for the interacting many-body system of highly excited atomic nuclei, was later discovered in a variety of situations including single-particle systems with disorder or chaos. A fascinating theme in this context is the emergence of universal laws for the fluctuations of energy spectra and transport observables. After an introduction to the basic phenomenology, the talk highlights the role of symmetries for universality, in particular the correspondence between symmetry classes and symmetric spaces that led to a classification scheme dubbed the ``Tenfold Way''. Perhaps surprisingly, the same scheme has turned out to organize also the world of topological insulators.
\end{abstract}

\maketitle

Let me begin by expressing that I feel greatly honored to be this year's recipient of the Max-Planck medal. I also appreciate the opportunity to give a talk on some of the work that may have earned me this distinction.

To set the stage and give you a flavor of what is to come, let me remind you of the old but still fascinating story of universal conductance fluctuations (UCF). Predicted theoretically in the middle of the 1980s by Altshuler \cite{Altshuler} and by Lee and Stone \cite{LeeStone}, UCF was investigated in a large number of experiments. It was found that in a great variety of different mesoscopic systems -- such as a small gold ring for example, or an even smaller silicon MOSFET -- the electrical conductance displays characteristic fluctuations of the order of one when expressed in units of the conductance quantum $e^2/h$ (Fig.\ 1). What is most remarkable is that the size of the fluctuations in a broad range of parameters does \emph{not} depend on the system dimension, the disorder strength, etc., but only on a few fundamental symmetries.

It was realized early on that there exists a close connection with the fluctuations that had been observed decades earlier in the scattering cross section of slow neutrons on atomic nuclei. This far reaching connection is at the very root of what I have to say. It led, among other things, to the development of a broad framework in which to model and calculate mesoscopic effects such as UCF.

\section{Symmetries}

Instead of giving a table of contents, let me just say that the organizational plan of the talk is to explain the words in my title, the first of which is \emph{symmetries}.

Let us start at the very beginning and recall what is meant by a symmetry in the context of quantum mechanics. Following Wigner (who was awarded the 1963 Physics Nobel Prize for his foundational work on symmetry principles and their application to nuclear and particle physics) a symmetry in quantum mechanics is primarily a transformation on the rays of Hilbert space with the property that all transition probabilities are preserved.

Now one corner stone of our subject is a theorem (attributed to Wigner) stating that any quantum mechanical symmetry lifts to Hilbert space as a linear operator which is either unitary, or anti-unitary. In the former case the Hermitian scalar product of Hilbert space is preserved, in the latter case it is preserved up to complex conjugation. I wish to make two remarks here.

First comes the obvious statement that symmetries always form a group, $G\,$. Indeed, if two operators $g_1$ and $g_2$ are symmetries, then so is their composition $g_1 g_2$.

The second remark is that in order for an operator $g$ to be a symmetry of a quantum system with Hamiltonian $H$, we require that $g$ commutes with $H$. Thus so-called chiral symmetries, which \emph{anti}-commute with the Hamiltonian -- a prominent example is the chirality operator $\gamma_5$ which anti-commutes with the massless Dirac operator -- do \emph{not} qualify as symmetries, at least not in the sense of this talk. (I must emphasize this point because my work on symmetry classification is often cited in an abridged form that overlooks or ignores this aspect.)

Now, why should we care about any symmetries? 
We all know, of course, that symmetries in spectroscopy lead to selection rules and relations among the transition rates; and in the case of integrable systems they are the key feature responsible for integrability. However, here our interest is in the exact opposite, namely disorder and quantum chaos. Let me give two examples to illustrate that symmetries do matter even in that case.

\subsection{Universal conductance fluctuations}

First, I would like to return to the theme of universal conductance fluctuations and focus on its symmetry aspects, referring to the cartoon in Fig. 2a for illustration.

By assuming the single-electron approximation and using that the electrical conductance $C$ is essentially a sum of squares of transmission amplitudes, we can make a picture of it as a sum over semi-classical paths crossing the disordered metal. Taking UCF to be a statement about the \emph{variance} of the fluctuations, we consider two pairs of paths for the transmission amplitudes and their conjugates. We also arrange them in a way that makes our graph phase-insensitive and thus immune to the process of disorder averaging.

The leading contribution to the variance is due to a form of `promiscuity' of our electron paths: after passing from the leads into the disordered region, the electron paths exchange partners and diffuse with the alien conjugate for a while before they rejoin their proper mate in preparation for exit. To compute the variance, we integrate over the promiscuous diffusion legs and the two positions of adultery and re-marriage. This amounts to doing a momentum integral involving two diffusion propagators $(q^2)^{-1}$, and as is well known, the result is of the order of unity, independent of the system size $L$ and other system-specific features:
\begin{displaymath}
 (e^2 / h)^{-2}\; \mathrm{var}\big( C(L) \big) \; = \; L^{d-4} \int \frac{d^d q}{(q^2)^2} \; \propto \; \mathrm{O}(1) \,.
\end{displaymath}

To pinpoint the role of symmetries, let us review a little bit of theory. On very elementary grounds, the Green operator $G(z) = (z-H)^{-1}$ satisfies the resolvent identity $(z-w)\, G(z)\, G(w) = G(w) - G(z)$. Now if our single-electron Hamiltonian is Hermitian and conserves particle number (a property that can be viewed as a consequence of a global $\mathrm{U}(1)$ gauge symmetry) then the complex conjugate of the retarded Green's function is equal to the advanced Green's function with the initial and final states exchanged: $\overline{G_{a\,b}^+} = G_{b\,a}^-$. By closure and the resolvent identity, this relation results in a sum rule for the sum of squares:
\begin{displaymath}
 \sum\nolimits_b \big\vert G_{a\,b}^+ \big\vert^2 = (2\mathrm{i}\varepsilon)^{-1} (G_{a\,a}^- - G_{a\,a}^+) ,
\end{displaymath}
and the Fourier transform of the disorder-averaged squared amplitude has the form of a massless diffusion propagator. Thus we see that the $\mathrm{U}(1)$ symmetry of particle number conservation gives rise to a massless mode, namely the ``diffuson''.

What happens if the system has more symmetries? Well, then we should expect more massless modes! For example, if magnetic fields are absent and the disordered system is invariant under the operation of reversing the time direction, then the Green's function remains the same when we exchange the initial state with the time-reversed final state. By the line of reasoning of before, this symmetry relation gives rise to another sum rule and hence another massless mode called the ``cooperon''.

Returning to our graphical illustration, in the presence of time-reversal symmetry there exists a second type of graph (Fig.\ 2b), which is different from the earlier one but still immune to phase cancelations due to disorder averaging. In this UCF graph, the promiscuous legs are cooperons, as the two arrows indicating the direction of motion point the opposite way, whereas previously they pointed the same way. Its numerical contribution to the variance is the same as that of the other graph featuring the diffuson. Thus the outcome is that the symmetry under time reversal makes the variance double.

\subsection{Symplectic wires}

Let me move on to a second example, namely that of disordered `symplectic' wires, meaning electrons in a quasi-one dimensional geometry with spin-orbit scattering and, again, time-reversal invariance. Here the symmetries of the disordered quantum system turned out to be of truly spectacular consequence.

For concreteness, let us consider a model of Dirac fermions in one space dimension with coordinate $x$. The model is built from $N$ right-moving modes (or channels) coupled by a Hermitian random matrix $A(x)$, $N$ left-moving channels coupled by the transpose of $A$, and a skew-symmetric random matrix $B(x)$ causing backscattering, i.e., coupling between the right- and left-movers:
\begin{displaymath}
 H = \begin{pmatrix} v_F \frac{\hbar}{\mathrm{i}} \frac{\partial} {\partial x} + A(x) & B(x) \cr {B^\dagger (x)}^{\vphantom{\dagger}} &
 - v_F \frac{\hbar}{\mathrm{i}} \frac{\partial} {\partial x} + A^T (x) \end{pmatrix} ,
\end{displaymath}
where $A^\dagger = A$ and $B^T = - B$. The particular form and arrangement of the random matrices $A$ and $B$ places this model in what I call ``class $A$II''. A hallmark of this class is Kramers degeneracy of the energy eigenvalues.

Now, for Gaussian random variables and large channel number $N$, the disorder averaged observables of this model can be computed by the mapping to a nonlinear sigma model due to Wegner and Efetov. In a Letter published in 1992 \cite{MRZ-PRL1992}, I solved the sigma model to derive an exact expression for the average electrical conductance $\langle C \rangle$ of a wire of length $L$. The striking feature of my result, which was controversial (and in fact considered to be unphysical) at that time, is that the conductance does not go to zero with increasing length as one might have expected from the standard scenario of exponential localization in one dimension, but approaches a constant ($1/2$ in units of $e^2/h$).

One decade later a Japanese colleague \cite{Takane2004} made perfect sense of this result by pointing out that the large-$N$ limit I had computed was a mixture of contributions from even and odd channel number $N$, and that one really should separate even from odd. Thus the good way of presenting and interpreting my result is in a split form:
\begin{align*}
 &(e^2 / h)^{-1} \langle C \rangle \; = \; {\textstyle{\frac{1}{2}}} f_0(L) + {\textstyle{\frac{1}{2}}} f_1(L) \cr &\text{even\;} N : \quad f_0(L) = \frac{32}{9} \left( \frac{\pi \xi}{2L} \right)^{3/2} \!\!\!\! \mathrm{e}^{-L/2\xi} + ... \cr &\text{odd \;} N : \quad
 f_1(L) = 1 \; + \; 2 \, \mathrm{e}^{-4L/\xi} + ...
\end{align*}
where for brevity I am not showing the exact result but just the leading behavior for a long wire. For even $N$ exponential localization does hold, but for odd $N$ there is one perfectly conducting channel (or zero mode).

Let me emphasize that the existence of this mode is \emph{not} an artifact of the Dirac approximation used, or the limit of large channel number, or any other specific feature. Rather, it is a universal and inevitable consequence of time-reversal symmetry for spin-1/2 electrons in combination with the channel number being odd. As a matter of fact, the zero mode that showed up in my super-Fourier analysis is nothing but the celebrated edge state of the quantum spin Hall insulator!

We nowadays understand that this mode is robust on stringent topological grounds and, in particular, is stable with respect to disorder. (Sadly, I did not know about the topological significance of odd channel number at the time when my work was done.)

\section{Symmetry Classes}

So much for symmetries and how in disordered systems they give rise to massless modes with observable consequences for transport. On to the next word in my title: \emph{symmetry classes}. Here I must warn you that I need to be moving rapidly and will be rather brief. For a more comprehensive and leisurely account of this story, you may want to consult my web pages where you will find a colloquium style talk with slides and text included.

In short, by ``symmetry class'' I mean an organizational principle which is relatively new and was \emph{not} part of the vocabulary of mesoscopic physics fifteen years ago. At that time, one spoke of \emph{universality classes}.

To establish the physical context, let me recall a striking statement referred to as the random-matrix universality conjecture for spectral fluctuations. It is this: take any of the linear equations of wave mechanics, be it the Schr\"odinger equation, the Dirac equation or, for that matter, the wave equation of Maxwell electrodynamics, and look at the spectrum of energy levels. If there is enough disorder or chaos, then you will see level fluctuations which obey the laws predicted by random matrix theory for the appropriate symmetry class and in a certain limit called the ergodic regime. This finding has been made in numerous examples, beginning with the neutron resonances observed in compound nucleus scattering.

To single out one example, numerical diagonalization shows that the energy levels of the Sinai billiard, a quantum chaotic system with two degrees of freedom, exhibit the level correlations of the so-called Gaussian Orthogonal Ensemble, i.e., a real symmetric random matrix. This example was of great importance for the historical development, as it led Bohigas, Giannoni, and Schmit in 1984 to state the said universality conjecture. \cite{BGS1984}

The classification principle behind this kind of universality had been formulated some 20 twenty years earlier by Dyson in a famous paper \cite{Dyson3fold} called ``Threefold Way''. Dyson proved that, in his own words, ``the most general kind of matrix ensemble, defined with a symmetry group which may be completely arbitrary, reduces to a direct product of independent irreducible ensembles each of which belongs to one of the three known types.''

Dyson had asked the following question. Given a finite-dimensional Hilbert space and a general symmetry group acting on it, how do the Hamiltonians look that commute with all the symmetries? His answer for the irreducible blocks was that there exist but 3 possibilities: the matrix of the Hamiltonian is either complex Hermitian, or real symmetric, or has entries that are quaternions.

It should be emphasized that this organization by what we nowadays call the Wigner-Dyson symmetry classes is very coarse and relies on nothing but linear algebra. In fact, a symmetry class is simply a type of real vector space, or linear space, for the Hamiltonian matrix to be in. It is only by putting probability measures on these matrix spaces -- the simplest ones are the so-called Gaussian measures or Gaussian ensembles -- and analyzing the statistical correlations that one discovers the universality mentioned earlier. Thus the notion of \emph{symmetry class} is a notion more basic and more primitive than the notion of \emph{universality class}.

\subsection{The need for an extended scheme}

Now around the middle of the 1990s, there was an accumulation of evidence that some extension or variant of Dyson's threefold way was called for. First of all, it was found that the eigenvalue statistics of the massless Dirac operator which, for example, is relevant for quantum chromodynamics, could be modeled by so-called chiral ensembles of a type beyond Dyson's classification. \cite{ChiralRMT} Second, in mesoscopic physics the process of Andreev reflection at the boundary between a normal metal and a superconductor was found to give rise to quantum interference phenomena not present in the Wigner-Dyson symmetry classes. \cite{AZ1997} Third, mathematicians had placed the Riemann zeta function in an ensemble of similar functions and found statistical phenomena which did not fit into the Wigner-Dyson scheme. \cite{KatzSarnak}

Let me single out the case of superconductors for a little bit of detail. The theoretical background here is that the Gorkov-Green operator satisfies a set of relations (even when there are no physical symmetries at all) which are just due to the canonical anti-commutation relations for fermions. By similar reasoning as before, these relations give us a sum rule which in turn gives rise to what Altland and I called the $D$-type diffuson. \cite{AZ1997} The `mass' of this mode goes to zero as the energy of the quasi-particle approaches the chemical potential, and its strength is given by the low-energy density of states.

Let us come back, once more and now for the final time, to the subject of universal conductance fluctuations and our cartoon, featuring this time the $D$-type diffuson (Fig.\ 1c). The novel feature here is that the promiscuous legs are formed by pairing retarded with retarded and advanced with advanced Green's functions.

To achieve experimental verification of the $D$-type diffuson mode, it is not a good idea to look for its effects in the transport of electric charge, which is shunted in a superconductor by the condensate. However, for a spin-singlet superconductor the effect can be detected by measuring the spin transport, and if the condensate carries spin, then one can still look at the heat transport due to the quasi-particles.

\subsection{The Tenfold Way}

After this example, let us get on to the systematics. Based on the classification of massless modes, Altland and myself argued that in addition to the 3 classes of Wigner-Dyson and the 3 chiral classes of Verbaarschot, \cite{Verbaar3fold} there were 4 classes which can be realized in superconductors, or in metals in proximity with superconductors, thus raising the count to a grand total of ten. We also claimed that this was it, and ``no further classes will be found''. Later, in joint work with two colleagues from Bochum mathematics, this conjecture was given a precise formulation and proved. Here is an outline.

First we have to specify the rules of the game. We refine the setting of the Threefold Way by replacing Dyson's general Hilbert space by the more elaborate structure of a Fock space for fermions. Following Dyson, we adopt the setting of a symmetry group acting on Fock space by unitary and anti-unitary operators. We require the group of unitary symmetries to be defined on the single-particle space $V$ and extend it to Fock space $F$ in the natural way. (This requirement, I should say, excludes Yangian and other quantum group symmetries, which arise at the many-particle level.) There is no further restriction; in Dyson's words: the group of unitary symmetries may be completely arbitrary.

As for the anti-unitaries, we allow for the presence of time-reversal symmetry, which is defined on the single-particle Hilbert space and extends to Fock space in the usual way. Moreover, the structure of Fock space opens the possibility for another anti-unitary operation to be a symmetry; this is particle-hole conjugation, transforming the particle vacuum into the fully occupied state, and in general, a state of $n$ particles into a state of $n$ holes.

We can now formulate the problem to be solved. Let there be any
fermionic Fock space carrying a $G$-action, where $G$ is an arbitrary
symmetry group made from generators as described above; i.e., with
the most general element $g \in G$ being any combination of unitary
symmetry operations and/or time reversal and/or particle-hole
conjugation. Our object of interest is the set $\mathcal{H}$ of all Hamiltonians which are $G$-invariant one-body operators, i.e., operators that are \emph{quadratic} in the particle creation and annihilation operators and \emph{commute} with all operators from the symmetry group $G$.

The question then is: what can we say about the structure of the set
$\mathcal{H}$? Can we classify the types of irreducible block which
occur in this setting?

After toying with this question for a number of years, the final and
definitive answer was given in a paper with Heinzner and Huckleberry: \cite{HHZ} every irreducible block which occurs in this setting corresponds a classical irreducible symmetric space, and conversely, every classical irreducible symmetric space arises in this way. (More precisely, the time-evolution operators are in a compact symmetric space, while the Hamiltonian are in a Euclidean symmetric space that arises by linearization.)

Before explaining what is meant by a symmetric space, let us do a quick example, namely that of class $D$, where one has no symmetries at all. If we introduce a basis of Majorana operators $\psi_{2l-1} = (c_l + c_l^\dagger)/\sqrt{2}$ and $\psi_{2l} = (c_l - c_l^\dagger) / \sqrt{2} \mathrm{i}$ ($l = 1, \ldots, N$), we can express our general one-body Hamiltonian in the simple form
\begin{displaymath}
    H = \sum_{1 \leq a < b \leq 2N} X_{ab} \, \psi_a \psi_b ,
\end{displaymath}
with imaginary coefficients $X_{ab}$ that constitute a skew-symmetric matrix. By exponentiating the Hamiltonian one gets time evolution operators $\mathrm{e}^{-\mathrm{i}tH/\hbar}$ in the even-dimensional orthogonal group, $\mathrm{SO}(2N)$, which is called a symmetric space of type $D$ (hence the name class $D$).

Experimental realizations of this class occur in disordered superconductors with spin-triplet pairing and $T$-breaking $p$-wave order -- a much debated candidate is $\mathrm {Sr}_2 \, \mathrm{Ru}\, \mathrm{O}_4$. A non-charged realization is the $A$-phase of ${}^{3} \mathrm{He}$. Moreover, a number of other realizations have recently been proposed, \cite{Majoranas} motivated by the far goal of engineering a topological quantum computer.

\section{Symmetric spaces}

Finally, to give some meaning to the statement of our theorem, let me explain the last word of the title, namely \emph{symmetric spaces}. Recall that in Riemannian geometry there exists an object called the Riemann curvature tensor. By definition, a symmetric space is a Riemannian manifold $X = G/K$ with a Riemann curvature tensor which is covariantly constant: $\nabla R = 0$.

The simplest example of a such a space is the round two-sphere $X = \mathrm{S}^2$ with line element induced by the Euclidean distance of three-dimensional space; in spherical polar coordinates this is $ds^2 = d\theta^2 + \sin^2 \theta\, d\phi^2$.

Another example is the set $X$ of all complex $n$-dimensional subspaces in $m+n$ dimensions. Such a space is called a Grassmann manifold, $\mathrm{Gr}_n (\mathbb{C}^{m+n})$, or Grassmannian for short. Note that any two $n$-dimensional subspaces can be mapped into each other by a unitary transformation of the full space. Since nothing changes when we just transform within a fixed subspace and its orthogonal complement, one has the identification $X = \mathrm{Gr}_n(\mathbb{C}^{m+n}) = \mathrm{U} (m+n) / \mathrm{U}(m) \times \mathrm{U}(n)$ of the Grassmannian with a quotient of unitary groups.

Here are a few facts about symmetric spaces. For one, they were completely classified
%
%
by the French geometer Elie Cartan. Apart from a finite number of exceptional spaces, they come in 10 large families, which Cartan called $A$, $A$I, $A$II, $A$III, $BD$, $BD$I, $C$, $C$I, $C$II, and $D$III.
For another fact, let me note that any $G$-invariant symmetric $(0,2)$-tensor on a symmetric space $G/K$ must be a constant multiple of the metric tensor.

There exist several places in physics where symmetric spaces appear.
For one, they are natural candidates for order parameter spaces. (Indeed, when the symmetry group $G$ of a physical system spontaneously breaks down to a subgroup $K$, the order parameter space is the quotient $G/K$.) For another, symmetric spaces serve as target spaces for a certain type of field theory, the so-called nonlinear sigma models. These models have the important property of being one-parameter renormalizable in two dimensions, i.e., under a change of short-distance cutoff of the field theory the metric tensor of the symmetric space (defining the nonlinear sigma model) changes only by a multiplicative constant. This follows from the mentioned fact that the metric is the only $G$-invariant tensor of its kind. Third, I have already spoken about the correspondence between symmetric spaces and symmetry classes. And last but not least, many-fermion ground states in the mean-field approximation of Hartree-Fock-Bogoliubov organize into symmetric spaces.

Let me now elaborate on the last point, as this will lead us to the final theme of the talk: topological insulators.

\subsection{Mean-field ground states}

As before, let $c_l^\dagger$ denote the Fock operator creating a single particle in the state labeled by $l$, and let $c_l$ be the corresponding annihilation operator. The Fock vacuum is uniquely characterized by the property of being annihilated by all of the annihilation operators: $c_l \, \vert \mathrm{vac} \rangle = 0$ ($l = 1 , 2, \ldots$). Now define the notion of \emph{quasi-particle vacuum} $\vert \widetilde{ \mathrm{vac}} \rangle$ by an analogous property:
\begin{displaymath}
 \widetilde{c}_l \, \vert \widetilde{\mathrm{vac}} \rangle = 0 \qquad (l = 1 , 2, \ldots),
\end{displaymath}
where the new operators $\widetilde{c}$ are the result of making a Bogoliubov transformation: $\widetilde{c}_l = \sum_{l^\prime} ( c_{ l^\prime} \, u_{l^\prime l} + c_{l^\prime}^\dagger \, v_{l^\prime l})$. Such vacua are also referred to as many-body ground states in the Hartree-Fock-Bogoliubov mean-field approximation. Note in particular that by specializing to
\begin{displaymath}
 \widetilde{c}_l^{\vphantom{\dagger}} = c_l^\dagger\quad (1 \leq l \leq N),\quad \widetilde{c}_l = c_l \quad (l > N),
\end{displaymath}
we get an $N$-particle Slater determinant state.

In the presence of a group $G$ of symmetries we require the quasi-particle vacuum state to be invariant under the $G$-action. (The context determines whether or not this is a reasonable condition to impose. In any case, it is part of the physical setting and mathematical statement I am driving at.) Let us now fix some symmetry group $G$, but keep the parameters of the system variable. As the mean-field Hamiltonian varies, so does its ground state, the quasi-particle vacuum $\vert \widetilde{\mathrm{vac}} \rangle$. Omitting the irrelevant overall phase factor, we introduce the notation $x \equiv \mathcal{R}\, \vert \widetilde{ \mathrm{vac}} \rangle$ for the ray of the ground state.

The space of quasi-particle vacua can be made into a Riemannian manifold $X$ by taking the geodesic distance between two such vacua $x_1$ and $x_2$ to be the positive number $t$ if the ray of one is obtained from the ray of the other by applying the exponential of an anti-Hermitian particle-hole type operator $A$ of norm $t$:
\begin{displaymath}
 \mathrm{dist}(x_1 , x_2) = t \; \Longleftrightarrow \; \mathcal{R}\, \vert {\widetilde{\mathrm{vac}}}_2 \rangle = \mathcal{R}\, \mathrm{e}^{A} \vert {\widetilde{\mathrm{vac}}}_1 \rangle , \;\; \parallel \! A \! \parallel \; = t .
\end{displaymath}

For example, let us take $G$ to be the $\mathrm{U}(1)$ symmetry group underlying conservation of particle number. For that choice, quasi-particle vacua are $N$-particle Slater determinants, and the Riemannian manifold of such states is a Grassmannian, $\mathrm{Gr}_N(V)$.  Indeed, an $N$-particle Slater determinant is completely defined by specifying just its $N$-dimensional subspace of occupied states inside the single-particle Hilbert space $V$. Moreover, the geometry of $\mathrm {Gr}_N (V)$ leads to the geodesic distance above.

One may now pose the question: what can be said about the structure of $X$ in general? Well, as a corollary of the theorem proved in \cite{HHZ}, the answer is that for any symmetry group $G$ the manifold of $G$-invariant quasi-particle vacua is a symmetric space (or more precisely: a product of classical irreducible symmetric spaces). This is the punch line of my talk.

\subsection{Quantum Spin Hall Insulator ($d = 2$)}

Now, to make a connection with some of the physics discussed at this March meeting, let us see how the quantum spin Hall insulator emerges from our picture.

We again assume that particle number is conserved and take the symmetry group $G$ to be generated by time reversal (for spin-1/2 electrons) and by a group $\Gamma$ of translations. The Fourier dual of the translation group, namely the (first) Brillouin zone, is denoted by $\widehat{\Gamma}$.

In the two-dimensional case at hand, our quasi-particle vacua are Hartree-Fock mean-field ground states. Because momentum $k$ is conserved due to translation symmetry, we may describe these ground states by assigning to each value $k \in \widehat{\Gamma}$ the corresponding vector space, $V(k)$, of occupied states (or, using the language of solid state physics, the vector space spanned by the valence band states at momentum $k$). If the system is a band insulator, i.e., the Fermi energy sits in a gap, this assignment $k \mapsto V(k)$ depends continuously on $k$. It is an example of what is called a ``vector bundle'' in mathematics.

In the presence of time-reversal symmetry we require that the time-reversal operator $T$ applied to a valence state of momentum $k$ yields a valence state at momentum $-k$. For the special case of a $T$-invariant momentum $k_0 = - k_0$ it follows that the vector space of valence states is $T$-invariant: $T\, V(k_0) = V(k_0)$. By the properties of the anti-unitary operator $T$ this implies that the vectors of $V(k_0)$ organize into so-called Kramers pairs.

Now there exists a mathematical tool called $K$-theory, which was developed in the 1950s (in the general context of the Atiyah-Singer index theorem) for the purpose of classifying vector bundles up to isomorphism. In our case (and for $\widehat{\Gamma} = \mathrm{S}^2$, the case of a fluid where $k_0 = 0, \infty$ are the only $T$-invariant momenta) $K$-theory tells us that there exist exactly two isomorphism classes of such vector bundles. Physically speaking, we nowadays say that the topology of the situation gives us two distinct phases: the trivial band insulator phase and the quantum spin Hall phase (and every one of our Hartree-Fock mean-field fluid ground states belongs to one of the two).

To distinguish between the two cases, one looks at a topological invariant that measures the twisting of the vector bundle -- following Kane and Mel\'e \cite{KaneMele} one associates with $V(k)$ a certain Pfaffian, and the argument of this Pfaffian winds either an even or an odd number of times along a circle through $k_0 = 0$ and $k_0 = \infty$ in $k$-space.

\subsection{Alternative view (symmetric spaces)}

You might now be confused if not irritated. I announced a story about symmetric spaces, but then I turned it into a story about vector bundles. Well, I am sorry, but unfortunately it is not always possible to get a full understanding from just one narrow perspective.

So let us change our perspective now and go to an alternative viewpoint leaning on symmetric spaces. Recall that a choice of $n$-dimensional subspace $V(k) \subset \mathbb{C}^{m+n}$ determines a point $x \in X = \mathrm{U}(m+n)/\mathrm{U}(m) \times \mathrm{U}(n)$. In the vector bundle picture we associated with each value $k \in \widehat{\Gamma}$ a vector space $V(k)$, but we may equivalently associate with each $k$-value the corresponding point in $X$. Thus instead of a vector bundle we get a mapping, say $\psi : \widehat{\Gamma} \to X$. The condition $T V(k) = V(-k)$ due to time-reversal symmetry translates into a condition
$\widetilde{T} \psi(k) = \psi(-k)$. At $T$-invariant points $k_0$ this condition constrains $\psi$ to take values in a symplectic Grassmannian:
\begin{displaymath}
 \psi(k_0) \in X_0 = \mathrm{Sp}(m+n) / \mathrm{Sp}(m) \times \mathrm{Sp}(n) \subset X .
\end{displaymath}
Here you have it; that's the symmetric space formulation.

Quite generally, in this mean-field picture, topological phases are classified by equivalence classes (so-called homotopy classes) of mappings $\psi$ from momentum space $\widehat{\Gamma}$ into a symmetric space $X$
subject to a condition
\begin{displaymath}
 g \cdot \psi(k) = \psi(g \cdot k) \quad (\text{for all\; } g \in G_\mathrm{red}\,),
\end{displaymath}
for every element of the reduced group of symmetries which do not commute with the translations.

In 2009 most if not all of the examples of topological insulators and superconductors known at that time were put into a ``Periodic Table'' by Ludwig and co-workers, \cite{PerTable} following Kitaev. \cite{Kitaev10fold} Related to a mathematical result called Bott periodicity, this Table seems to have caused quite a stir in the community; in any case, it has been reprinted in numerous articles including a major review by Hasan and Kane. \cite{HasanKane}

Let me finish with a few remarks on this. First, the Table, as wonderful as it is, does not give the complete story, as there exist symmetry groups of physical relevance which are not present in the scheme of~ \cite{PerTable}. Second, the $K$ theory formalism used to compute it may miss some of the finer points of topology. In fact some of the Table's zero entries are not zero in all possible cases. My final remark is that, to the best of my knowledge, there exists no mathematical proof of the Table up to now (but I believe that such will soon become available).

\section{Conclusion}

In this talk I tried to cover quite a lot of ground. Starting from the role of symmetries in systems with disorder or chaos, I introduced the notion of symmetry classes and symmetric spaces, and then described a one-to-one correspondence between these, the Tenfold Way. While conceived for the purpose of classifying disordered fermions, this scheme turned out to have the rewarding feature of being relevant for a topic of current interest, namely topological insulators.

\newpage
\begin{figure*}
    \begin{center}
        \epsfig{file=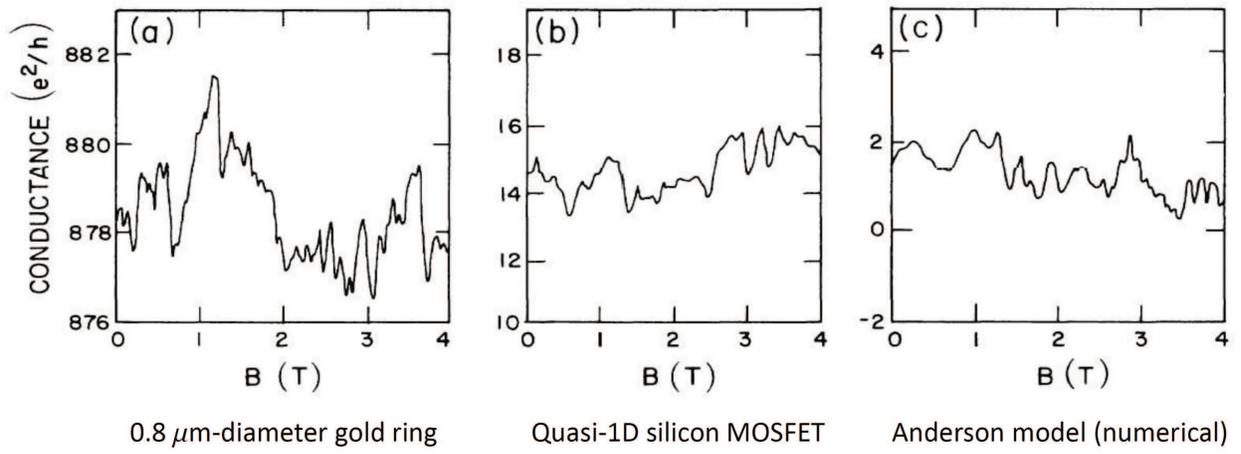,height=6cm}
    \end{center}
    \caption{Conductance fluctuations for various mesoscopic systems (taken from Lee, Stone, and Fukuyama, 1987).}
\end{figure*}

\begin{figure*}
    \begin{center}
        \epsfig{file=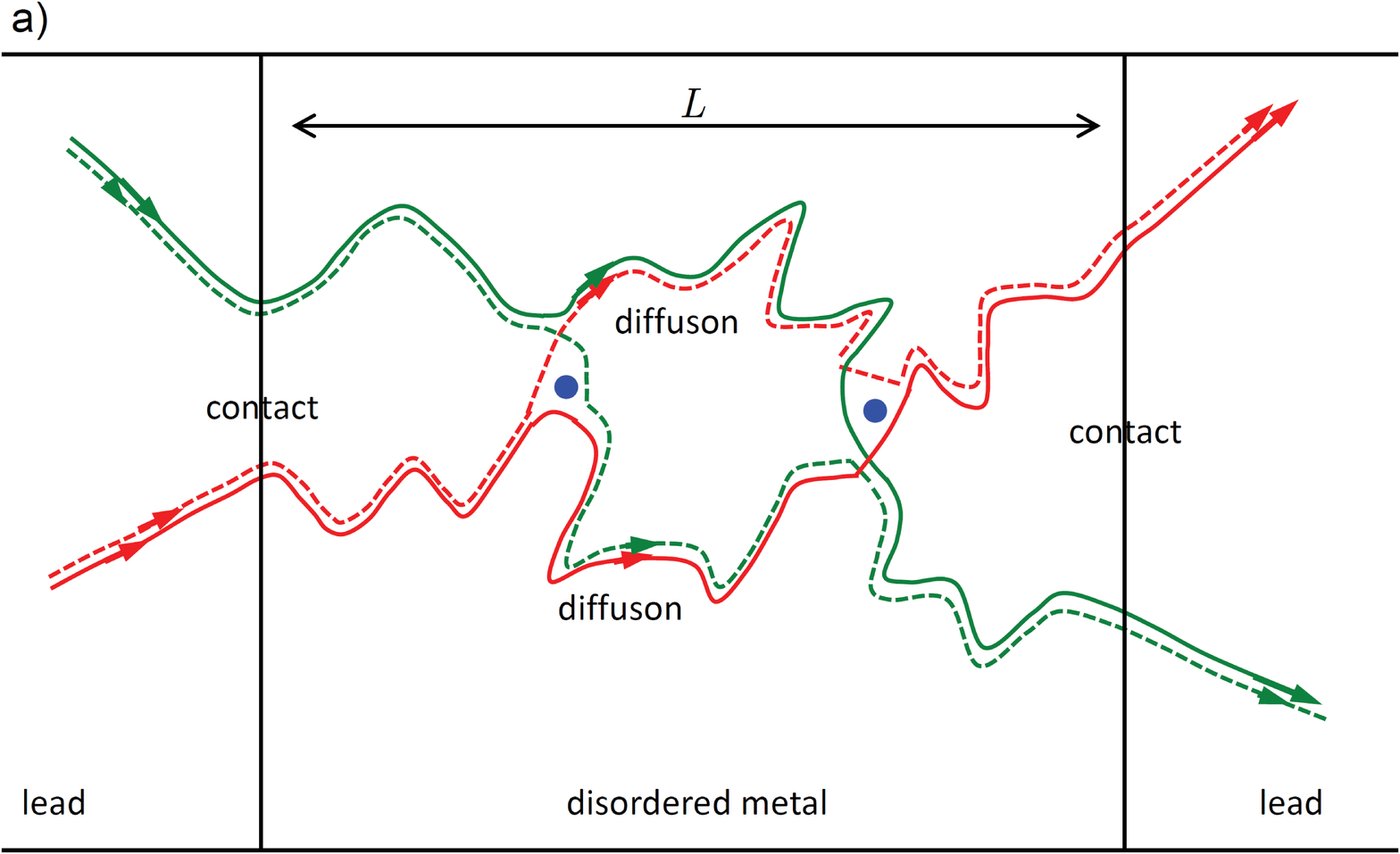,height=6cm}
    \end{center}
\begin{center}
        \epsfig{file=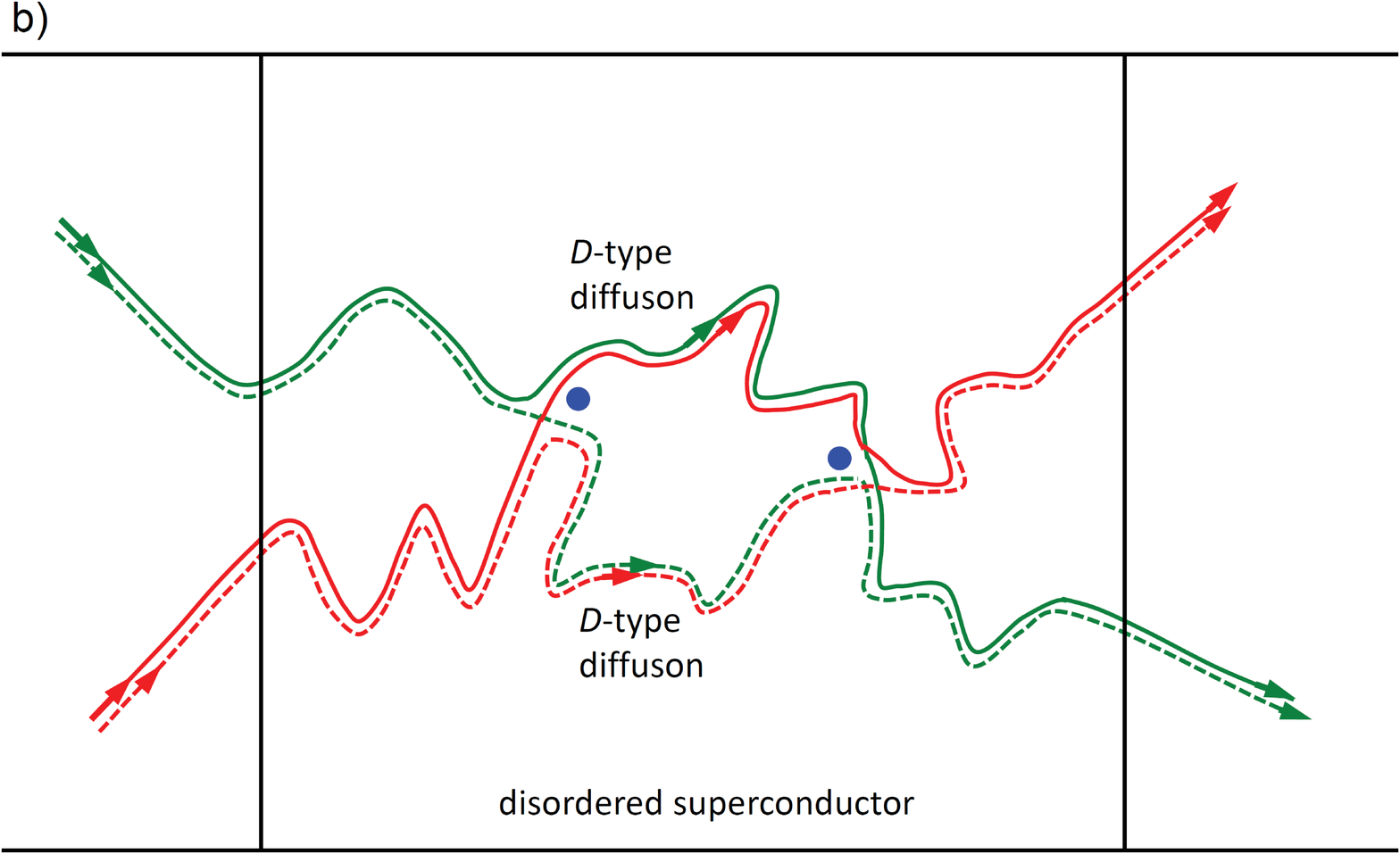,height=6cm}
    \end{center}
\begin{center}
        \epsfig{file=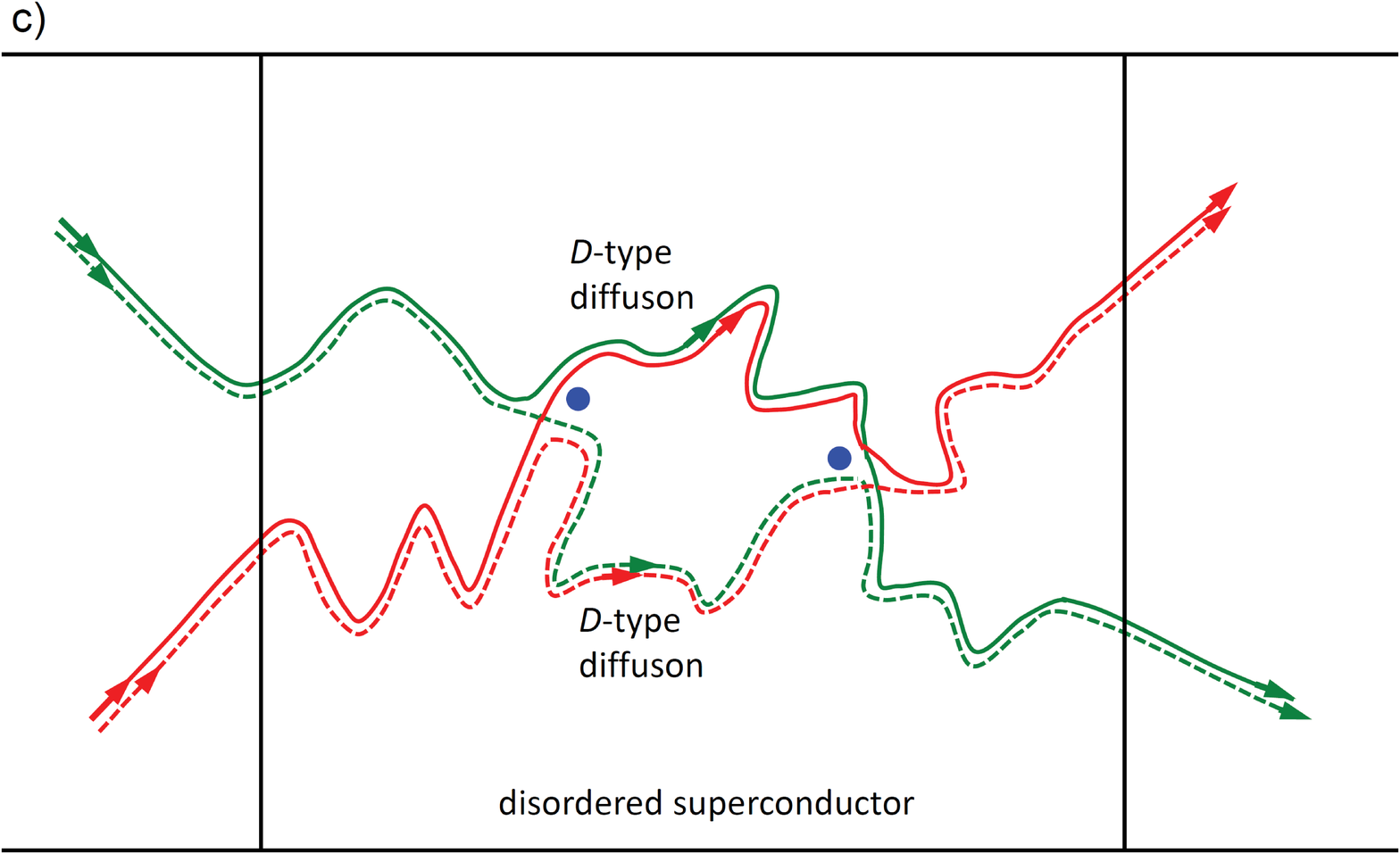,height=6cm}
\end{center}
    \caption{Semiclassical picture of contributions to the variance of the conductance. Color coding (red versus green) is used to distinguish between the paths for the two factors of the conductance entering the calculation of the variance. Full and dashed lines depict transmission amplitudes and their conjugates, respectively. a) Two-diffuson graph for a disordered metal. b) Two-cooperon graph. c) Graph built from two $D$-type diffusons in a disordered superconductor.}
\end{figure*}

\begin{thebibliography}{99}
\bibitem{Altshuler} B.L.\ Altshuler, JETP Letters {\bf 41} (1985) 648
%
\bibitem{LeeStone} P.A.\ Lee and A.D.\ Stone, Phys.\ Rev.\ Lett.\ {\bf 55} (1985) 1622
%
\bibitem{LeeStoneFukuyama} P.A.\ Lee, A.D.\ Stone, and H.\ Fukuyama, Phys.\ Rev.\ B {\bf 35} (1987) 1039
%
\bibitem{MRZ-PRL1992} M.R.\ Zirnbauer, Phys.\ Rev.\ Lett.\ {\bf 69} (1992) 1584
%
\bibitem{Takane2004} Y.\ Takane, J. Phys. Soc. Japan {\bf 73} (2004) 1430
%
\bibitem{BGS1984} O.\ Bohigas, M.J.\ Giannoni, and C.\ Schmit, Phys.\ Rev.\ Lett.\ {\bf 52} (1984) 1
%
\bibitem{Dyson3fold} F.J.\ Dyson, J.\ Math.\ Phys.\ {\bf 3} (1962) 1199
%
\bibitem{ChiralRMT} J.J.M.\ Verbaarschot and I.\ Zahed, Phys.\ Rev.\ Lett.\ {\bf 70} (1993) 3852
%
\bibitem{AZ1997} A.\ Altland and M.R.\ Zirnbauer, Phys.\ Rev.\ B {\bf 55} (1997) 1142
%
\bibitem{KatzSarnak} N.M.\ Katz and P.\ Sarnak, Bull.\ Amer.\ Math.\ Soc.\ {\bf 36} (1999) 1
%
\bibitem{Verbaar3fold} J.J.M.\ Verbaarschot, Phys.\ Rev.\ Lett.\ {\bf 72} (1994) 2531
%
\bibitem{HHZ} P.\ Heinzner, A.\ Huckleberry, and M.R.\ Zirnbauer {\bf 257} (2005) 725
%
\bibitem{Majoranas} See J.\ Alicea, arXiv:1202.1293, for an overview.
%
\bibitem{KaneMele} C.L.\ Kane and E.J.\ Mele, Phys.\ Rev.\ Lett.\ {\bf 95} (2005) 146802
%
\bibitem{PerTable} S.\ Ryu, A.P.\ Schnyder, A.\ Furusaki, and A.W.W.\ Ludwig, New J. Phys.\ {\bf 12} (2010) 065010
%
\bibitem{Kitaev10fold} A.\ Kitaev, AIP Conf.\ Proc.\ 1134 (2009) 22
%
\bibitem{HasanKane} M.Z.\ Hasan and C.L.\ Kane, Rev.\ Mod.\ Phys.\ {\bf 82} (2010) 3045
\end{thebibliography}
\end{document}